\documentclass[twoside]{dis04}
\def\runauthor{A. Kyrieleis}
\def\shorttitle{NLO photon impact factor}

\def\gs{{\gamma^*}}
\def\ep{{\epsilon}}

\def\br{{\bf r}}
\def\GGqqg{|\Gamma^{(0)}_{\gamma^*\to q{\bar q}g}|^2}
\def\Gqq{\Gamma_{\gamma^*\to q{\bar q}}}

\begin{document}

\title{The Real Corrections to the $\gamma^*$ impact factor: \\
First Numerical Results}

\author{Albrecht Kyrieleis\footnote{\uppercase{T}\lowercase{he work
presented in this talk was
done in collaboration with} \uppercase{J.B}\lowercase{artels},
\uppercase{U}\lowercase{niversity \uppercase{H}amburg}}}

\address{Department of Physics \& Astronomy, University of Manchester, \\
  Oxford Road, Manchester M13 9PL, U.K.\\
  E-mail: albrecht@theory.ph.man.ac.uk }

\maketitle

\abstracts{We have performed analytically the transverse momentum
integrations in the real corrections to the longitudinal $\gamma^*_L$
impact factor and carried out numerically the remaining
integrations. I outline the analytical integration and  present the
nume\-rical results: we have performed a numerical test and  computed those parts of the
impact factor that depend upon the energy scale $s_0$.}

\section{Introduction} 
The $\gs\gs$ total cross section is a
very attractive observable to be calculated in the framework of NLO BFKL.  To perform this
calculation and also to approach e.g. the problem of the color dipole
picture at NLO, the $\gs$ impact factor is needed at NLO. The NLO
corrections of  this impact factor are calculated from
the photon-Reggeon vertices for $q\bar q$ and $q\bar qg$ production,
respectively.  The virtual corrections 
involve the vertex  $\Gqq^{(1)}$ at one-loop level which has been calculated in
\cite{virt,kot}. As to the real
corrections, the squared vertex $\GGqqg$ is needed at tree level; it
has been computed in \cite{real,combine} (cf. also \cite{kot}). In \cite{combine} we have combined the
infrared  divergences of the virtual and of the real parts,
and we have demonstrated their cancellation.
What remains to complete the calculation of the NLO photon impact factor
are the integrations over the $q\bar q$ and $q\bar qg$ phase space,
respectively. 
A slightly different approach of calculating the NLO photon impact
factor has been proposed in \cite{FIK}. 

We have performed, for the case of longitudinal photon
polarisation, the phase space integration  in the real corrections \cite{BI}. 
After the  introduction of  Feynman parameters we carried out the
integration over transverse momenta analytically. 
This will allow for
further theoretical investigations of the photon impact
factor. In particular, the Mellin transform of the real corrections
w.r.t the Reggeon momentum can be calculated. This representation
(together with an analogous representation of the virtual corrections) 
 can  be a starting point for the resummation of the
next-to-leading logs(1/x) in the quark anomalous dimensions.

The procedure of arriving at finite NLO corrections to the $\gs$ impact
factor has been described in \cite{combine}: $\GGqqg$ has to be integrated over the $q\bar qg$ phase space. 
But there are two restrictions.
First, we have to exclude that region of phase space 
where the gluon is separated in rapidity from the $q\bar q$ pair (central
region); this configuration belongs to the LLA and has to be subtracted. 
To divide the $q\bar qg$ phase space an energy cutoff $s_0$ is
introduced which plays the role of the energy scale.
The virtual corrections to the impact factor are independent of $s_0$.
As a result, the integration of
the real corrections already allows to study the $s_0$ dependence 
of the NLO impact factor. Second, we need to take care of the infrared
infinities. The divergences in $\GGqqg$ due to the gluon
being either soft or collinear to either of the fermions are
regularised by subtracting the approximation of the squared vertex in
the corresponding limit. These expressions are then re-added and integrated
in $4-2\ep$ space-time dimensions, giving rise to poles in $\ep$,
which drop out in combination with the virtual corrections and to 
finite pieces. The subtraction of the collinear limit requires 
the introduction of a momentum cutoff parameter, $\Lambda$. The final  
NLO corrections must be independent of this auxiliary parameter.
In our numerical analysis we performed this important
test.
\begin{figure}[h]
\vspace{-.6cm}
\begin{center}
\centerline{\epsfxsize=1.5in\epsfbox{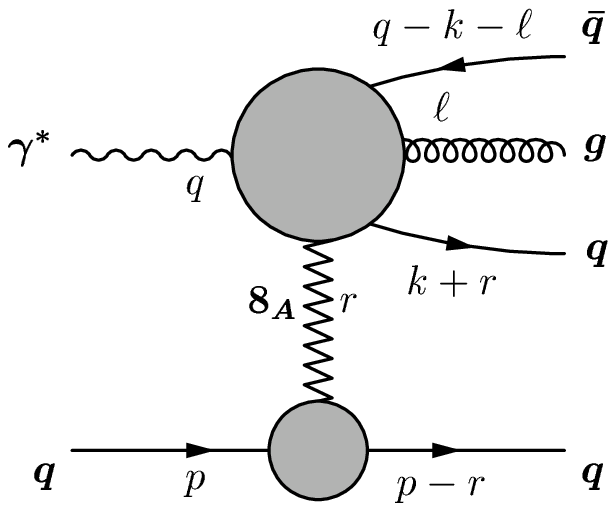}}
\caption[*]{\label{qqgkin}  The process $\gs+q'\to q{\bar
q}g+q'$ to calculate $\Gamma^{(0)}_{\gamma^*\to q{\bar q}g}$}  
\end{center}
\end{figure}
\vspace{-.7cm}
\section{The integration}
The $q{\bar q}g$ production vertex is calculated from
the process $\gs+q'\to q{\bar q}g+q'$ at tree level. The integration
of $\GGqqg$   therefore implies 
the integration of a sum of expressions, corresponding to products of Feynman diagrams
that differ in their denominator structure (for simplicity we will in
the following  use 'diagram' rather than  'product of diagrams'). In order to introduce Feynman parameters we therefore  split this sum and
treat each diagram (or pairs of them) independently. This
gives rise to divergences, that in the sum of all diagrams cancel
but show up in individual diagrams. 
The main task in the program of performing the integration analytically 
is the regularisation of these additional divergences in each individual
diagram. We used the subtraction method: From each divergent
diagram its approximation in the limit of divergence is subtracted and
re-added. The integrations in the re-added pieces are
carried out analytically in $4-2\ep$
space-time dimensions; the poles in $\ep$ are found to
cancel. 

As the  result of the regularisation and of the analytical integration
over transverse momenta we have obtained for each diagram (or pairs of
them) a convergent integral over the Feynman parameters and the
momentum fractions of the final state quark and gluon. We denote the sum of
these integrals from all diagrams (including the finite pieces from
the regularisation of the  additional divergences) by $\Phi_A$ and
$\Phi_F$ according to the separation  w.r.t. the color factors
 $C_A, C_F$. The integrand for each diagram
is a sum of terms corresponding to the subtractions necessary to 
regularise all divergences of the diagram (see \cite{BI}).
\section{Numerical results}
\begin{figure}[h]
\vspace{8cm}
\begin{center}
\includegraphics{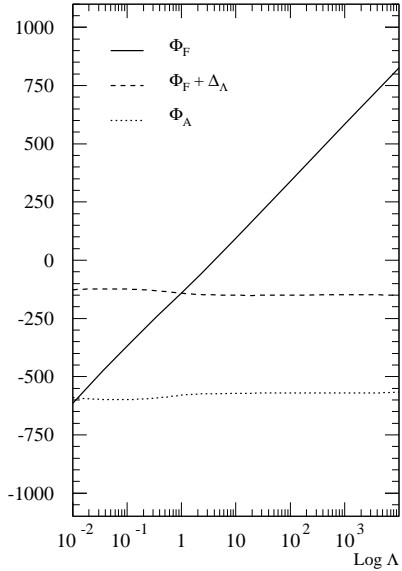}
\caption[*]{\label{lambda-dep} The dependence of the real corrections on
  $\Lambda$ (with  $\br^2 = s_0 = 1$) }
\vspace{-.9cm}
\end{center}
\end{figure}

We have carried out numerically the integrations over the Feynman parameters and
the momentum fractions of the quark and gluon for all
diagrams. 
As the result we have obtained values of the real
corrections,$\Phi^{\mathrm real}$, to the photon impact factor as a
function of $\Lambda/Q, s_0/Q^2$ and $\br^2/Q^2$, where $\br$ is the
transverse part of the Reggeon momentum and $Q^2$ is the photon
virtuality. In the following we suppress the scaling with $Q$ and
label the dimensionless variables $\Lambda, s_0,\br^2$. Besides $\Phi_A,
\Phi_F$ (see above) we include in $\Phi^{\mathrm real}$ those finite pieces from
the regularisation of the soft and collinear divergences that
depend upon  $\Lambda$ or  $s_0$ ($\Delta_{\Lambda}$ and
$\Delta_{s_0}$, respectively):
\begin{displaymath}
\Phi^{\mathrm real} = \,e^2 e_f^2 \;(\Phi_A + \Phi_F +
  \Delta_\Lambda + \Delta_{s_0})\:,\quad \Delta_\Lambda= -
  3\,C_F\,\frac{\Phi^{(0)}}{(2\pi)^2}\; \ln  \Lambda. 
\label{eq:realcorr}
\end{displaymath}
$\Phi^{(0)}$ denotes the LO $\gs$ impact factor, $ee_f$ is the  charge
of the quark. The full dependence
of the NLO impact factor on $\Lambda$ and $s_0$ is now included
in $\Phi^{\mathrm real}$. As said above, the NLO impact factor
and hence $\Phi^{\mathrm real}$ has to be independent of
$\Lambda$. Fig.\ref{lambda-dep} shows that, in fact, the $\Lambda$
dependence of $\Phi_A$ is very weak. As to the $C_F$ terms, $\Phi_F$ turns out to be 
proportional to $\ln \Lambda$. This growth with $\Lambda$, however, 
is fully compensated by $\Delta_\Lambda$. 
Note that $\Phi_F$ is a sum of many Feynman diagrams, 
whereas $\Delta_\Lambda \sim \ln \Lambda$. The compensation of
the $\Lambda$ dependence, therefore, represents a rather stringent test
of the calculation of the $\gs$ impact factor.

\begin{figure}[h]
\vspace{5.8cm}
\begin{center}
\includegraphics{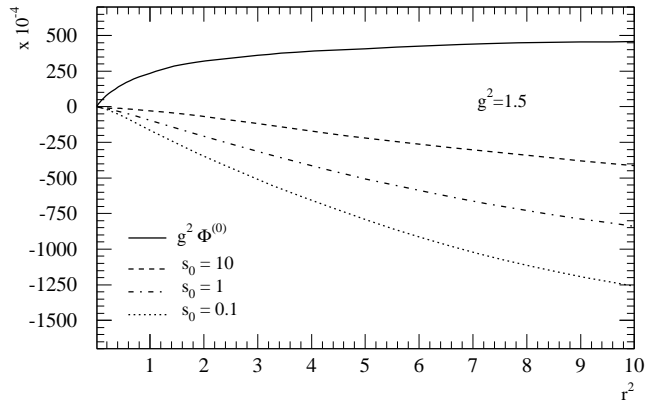}
\caption[*]{\label{ifac} $\Phi' $ at different different values of
$s_0$}
\vspace{-.9cm}
\end{center}
\end{figure}

Next we address the dependence of the NLO $\gs$ impact factor on the
energy scale $s_0$. Since, at the moment,
we only know  the real corrections, we compute, as a part of the full
LO and NLO impact factor: 
\begin{displaymath}
\Phi' = g^2\Phi^{(0)} + g^4\Phi^{\mathrm real}
\end{displaymath}
 where $g^2=4\pi \alpha_s$. We  set $e^2e_f^2=1$. For the photon virtuality we choose $Q^2=15$ GeV$^2$ which
leads $\alpha_s(Q^2)=0.18$ or $g^2=1.5$. Fig.\ref{ifac} compares
$\Phi'$ to the LO impact factor $g^2\Phi^{(0)}$ as function of
$\br^2$ at different values of $s_0$. 
The real corrections are negative and rather
large. More important, $\Phi'$ becomes, in absolute terms, more significant 
for smaller values of $s_0$. Since we
included all $s_0$ dependent terms in $\Phi'$, this implies that
the $\gs$ impact factor tends to become smaller with decreasing $s_0$. A physical
scattering amplitude (e.g. for the $\gs\gs$ scattering process), when
consistently evaluated in the framework of NLO BFKL has to be
invariant under changes of
$s_0$. Since a decrease of $s_0$ in the energy dependence
$(\frac{s}{s_0})^\omega$ will enhance the scattering amplitude, the combined
$s_0$ dependence of the impact factors and the BFKL Green's function
has to compensate this growth. Our result for the $s_0$ behaviour of
the $\gs$ impact factor is therefore, at least, consistent with the
general  expectation.

\end{document}